\documentclass[
 aps,
 amsmath,amssymb,
 twocolumn,
]{revtex4-2}

\usepackage[pdftex]{graphicx,color} % for publishing by pdflatex
\usepackage{comment}
\usepackage{natbib}
\usepackage{breakcites}
\usepackage{braket}
\usepackage{here}
\usepackage[normalem]{ulem}
\usepackage{hyperref}
\usepackage{url}
\usepackage{dcolumn}% Align table columns on decimal point
\usepackage{bm}% bold math

\begin{document}

\title{
Inverse Hamiltonian design  by automatic differentiation 
}

\author{Koji Inui}%\email{inui@aion.t.u-tokyo.ac.jp}
\author{Yukitoshi Motome}%\email{motome@ap.t.u-tokyo.ac.jp}
\affiliation{
 Department of Applied Physics, The University of Tokyo, Hongo, Tokyo 113-8656, Japan
}

\begin{abstract}
An ultimate goal of materials science is to deliver materials with desired properties at will. 
In the theoretical study, a standard approach consists of constructing a Hamiltonian based on phenomenology or first principles, calculating physical observables, and improving the Hamiltonian through feedback.
However, there is also an approach that bypasses such a cumbersome procedure, namely, to obtain an appropriate Hamiltonian directly from the desired properties. 
Solving the inverse problem has the potential to reach qualitatively different principles~\cite{Franceschetti1999,Zunger2018,MGI}, but most research to date has been limited to quantitative determination of parameters within known models~\cite{RevModPhys.83.943,Hart2005,PhysRevB.95.064407,ChiPhysLett.wei,PhysRevB.97.075114}.
Here, we present a general framework that enables the inverse design of Hamiltonians by optimizing numerous parameters using automatic differentiation. 
By applying it to the quantum anomalous Hall effect (AHE), we show that our framework can not only rediscover the Haldane model~\cite{PhysRevLett.61.2015} but also automatically generate a new Hamiltonian that exhibits a six-times larger AHE. 
In addition, the application to the photovoltaic effect (PVE)~\cite{PhysRev.134.A1313,doi:10.1063/1.1655453,PhysRevB.23.5590,PhysRevLett.109.236601} gives an optimal Hamiltonian for electrons moving on a noncoplanar spin texture, which can generate $\sim 900$~A/m$^2$ under solar radiation. 
This framework would accelerate materials exploration by establishing unprecedented models and principles beyond human intuition and empirical rules.
\end{abstract}

\maketitle

\section*{Main}
\label{main}

\begin{figure*}[htbp]
        \centering
        \includegraphics[width=18cm,pagebox=cropbox,trim=0 60 0 70, clip]{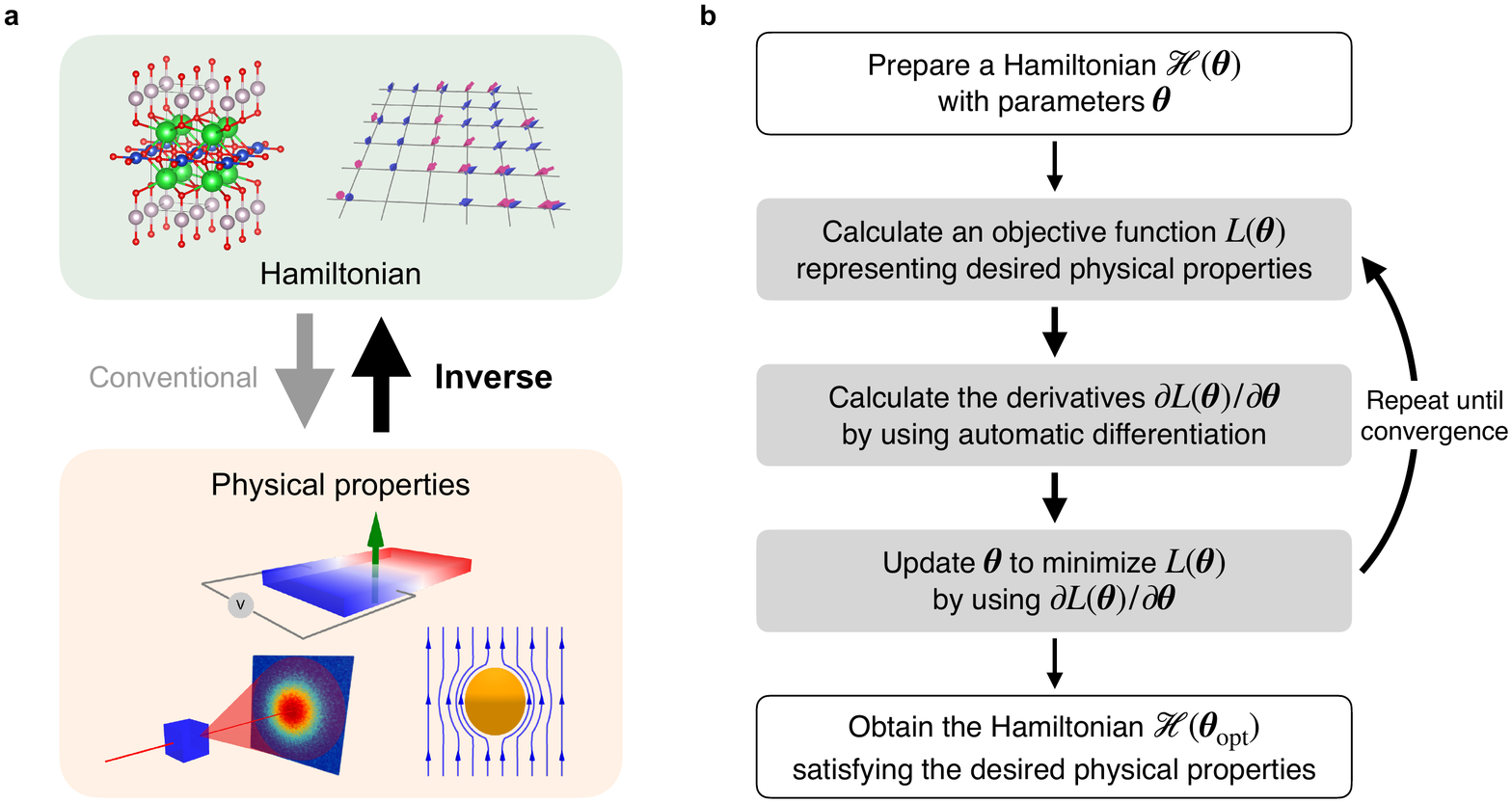} %<left> <lower> <right> <upper>
        \caption{
        {\bf Inverse design of Hamiltonian.} 
        {\bf a}, 
        In the conventional approach, the Hamiltonian is first constructed based on phenomenology or first principles, and then, the optimal parameters of the Hamiltonian are explored through physical properties calculated from the Hamiltonian. 
        In contrast, in the inverse approach, the desired physical properties are prepared first, and then, the Hamiltonian to realize them is obtained directly. 
        {\bf b}, 
        Flowchart proposed in the present study to solve the inverse problem by using automatic differentiation.
}
        \label{Fig:flow}
\end{figure*}

A conventional theoretical approach to materials exploration is to search for Hamiltonians that produce physical properties of interest (Fig.~\ref{Fig:flow}$\bf{a}$). 
This is not only tedious but also nontrivial since the parameter space to be explored is usually unknown a priori.
Therefore, most of the research to date has been conducted for the known Hamiltonians and their extensions. 
However, these approaches make it difficult to reach qualitatively new models and principles.
In contrast, the inverse approach to find appropriate Hamiltonians directly from the desired properties is not only efficient but also has the potential to unveil qualitatively new physics (Fig.~\ref{Fig:flow}$\bf{a}$). 
Over the past years, many proposals have been made in this direction~\cite{Franceschetti1999,Hart2005,Zunger2018,Long2021}, in particular, for estimation of the model parameters from experimental or simulation data~\cite{PhysRevB.95.064407,ChiPhysLett.wei,PhysRevB.97.075114} and construction of the Hamiltonians from the desired properties~\cite{Franceschetti1999,Hart2005,PhysRevResearch.3.013132,doi:10.1021/jp960518i,PhysRevLett.110.220503}. 
In most cases, however, the number of the parameters is a few due to the computational cost, and some methods have limitations on the models and physical quantities that can be applied. 
For these reasons, the previous research has also been limited to quantitative estimation of parameters within known Hamiltonians; 
it is still challenging to explore new models and principles by taking full advantage of the inverse problem.

To address these issues, we develop a framework that can automatically design a Hamiltonian with desired physical properties 
by using automatic differentiation. 
Automatic differentiation enables us to compute the analytic derivatives of any functions by adapting chain rules, which has been widely used in the field of deep learning in the process of backpropagation~\cite{Rumelhart1986}, even for over a trillion parameters~\cite{fedus2021switch}. 
The flowchart of our framework is shown in Fig.~\ref{Fig:flow}{\bf b}.
First, we prepare a Hamiltonian $\mathcal{H}(\bm{\theta})$ with a set of parameters $\bm{\theta}$. 
We also define the objective function $L(\bm{\theta})$ to be minimized for achieving the desired properties; for instance, if the objective is to maximize the expectation value of a physical quantity $P$, we can take $L({\bm\theta})=-\langle P({\bm\theta})\rangle$. 
Next, we compute the derivative $\frac{\partial L}{\partial {\bm{\theta}}}$ by automatic differentiation. 
Then, we update the Hamiltonian by changing the parameters $\bm{\bm{\theta}}$ according to $\frac{\partial L}{\partial {\bm{\theta}}}$. 
By repeating this procedure until $\bm{\theta}$ converge, we end up with the Hamiltonian $\mathcal{H}(\bm{\theta}_{\rm opt})$ that optimizes the desired properties, where $\bm{\theta}_{\rm opt}$ are the parameters after the convergence. 
The key aspect of this framework is that it enables us to optimize a large number of parameters simultaneously, which dramatically reduces the computational cost compared to the conventional approach.
More importantly, as there is in practice no limitation on the parameter space of the Hamiltonian, it may lead us to discover unprecedented Hamiltonians that are hard to obtain by human intuition. 
In the following, as a proof of concept of this framework, we show its application to two problems: AHE and PVE.

\begin{figure*}[htbp]
        \centering
        \includegraphics[width=18cm,pagebox=cropbox,trim=150 60 150 100, clip]{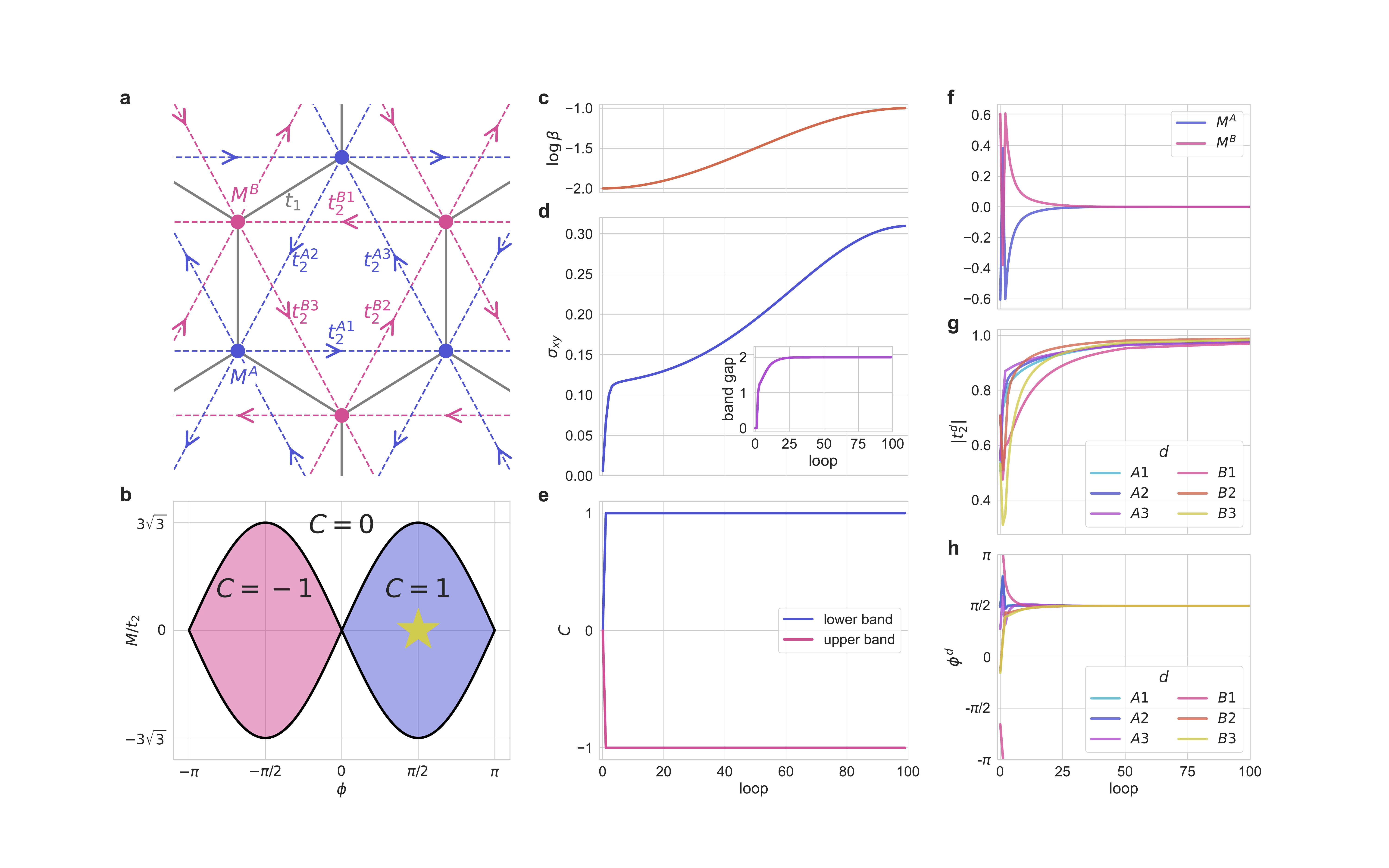} %<left> <lower> <right> <upper>
        \caption{
        {\bf Rediscovery of the Haldane model.} 
        $\bf{a}$, A tight-binding model on a honeycomb lattice in Eq.~\eqref{eq:H_honeycomb}. 
        There are 14 parameters including the amplitudes and phases of the second-neighbor hopping. 
        $\bf{b}$, Phase diagram of the Haldane model, where $M^A=+M$, $M^B=-M$, $t_1 = 1$, and $t_2^d=t_2\exp(i\phi)$.
        There are two topologically nontrivial phases with nonzero Chern numbers $C=\pm1$. 
        The yellow star represents where our framework reaches after the convergence. 
        $\bf{c}$,  Schedule of $\log \beta $, where $\beta$ is the inverse temperature. 
        $\bf{d, e}$,  Changes of the Hall conductivity $\sigma_{xy}$ ({\bf d}) and the Chern numbers $C$ for the two bands ({\bf e}) through the optimization process. 
        The inset in {\bf d} shows the change of the band gap. 
        {\bf f--h} , 
        Changes of the parameters: $M^a$ ({\bf f}), $|t_2^d|$  ({\bf g}), and $\phi^d$ ({\bf h}).
        }
        \label{Fig:haldane}
\end{figure*}

First, we demonstrate that our framework can automatically rediscover the Haldane model with a spontaneous quantum AHE~\cite{PhysRevLett.61.2015}. 
We consider a tight-binding model on a honeycomb lattice with two sublattices, whose Hamiltonian reads 
\begin{eqnarray}
\mathcal{H} = \sum_{i}M^a c_i^\dagger c_i + \sum_{\langle i,j \rangle} t_1 c_i^\dagger c_j+ \sum_{\langle \langle i,j \rangle \rangle} t_2^dc_i^\dagger c_j, 
\label{eq:H_honeycomb}
\end{eqnarray}
where $c_i^\dagger$ ($c_i$) is the creation (annihilation) operator of a spinless fermion at site $i$; 
the first term describes an on-site staggered potential with real coefficients $M^a$ ($a=A$ or $B$ denotes the sublattice), and the second and third terms represent the hopping of fermions to nearest- and second-neighbor sites, respectively. 
Here, we set $t_1=1$ and parametrize $t_2^d$ as $t^{d}_2 = \sigma(r^d)\exp({\rm i}\phi^d)$ with real variables $r^d$ and $\phi^d$, where $\sigma(x) = 1/(1+e^{-x})$ is a sigmoid function and $d$ denotes the direction of the second-neighbor hopping ($d=A1,A2,A3,B1,B2,B3$); 
see Fig.~\ref{Fig:haldane}{\bf a}. 
Thus, the model includes $14$ parameters in total represented as $\bm{\theta} = \{M^A, M^B, \{r^d\}, \{\phi^d\}\}$. 
The Haldane model is given by taking $M^A=+M$, $M^B=-M$, and $t_2^d=t_2\exp({\rm i}\phi)$ regardless of $d$.  
The phase diagram is shown in Fig.~\ref{Fig:haldane}{\bf b}, which has two topologically nontrivial phases with a spontaneous quantum AHE corresponding to the nonzero Chern numbers $C=\pm1$.

With this setup of $\mathcal{H}(\bm{\theta})$, we try to obtain a Hamiltonian that maximizes the AHE by the framework in Fig.~\ref{Fig:flow}{\bf b}. 
For this aim, we take the objective function as $L({\bm \theta}) = -\sigma_{xy}({\bm \theta})$, where $\sigma_{xy}$ is the Hall conductivity. 
Details of the calculations are described in Methods.
We find that $\sigma_{xy}$ increases monotonically through the optimization, as shown in Fig.~\ref{Fig:haldane}{\bf d}. 
Note that we introduce temperature and control it as shown in Fig.~\ref{Fig:haldane}{\bf c} to avoid that $\partial L / \partial \bm{\theta}$ becomes zero due to the quantization ($\beta$ is the inverse temperature).
In contrast to the continuous change of $\sigma_{xy}$, the Chern numbers of the two bands, which are separated by the band gap shown in the inset of Fig.~\ref{Fig:haldane}{\bf d}, converge quickly to $C\simeq\pm1$ in the very early stage of the optimization, as shown in Fig.~\ref{Fig:haldane}{\bf e}. 
The evolution of each parameter is plotted in Figs.~\ref{Fig:haldane}{\bf f}--{\bf h}. 
We find that both $M^A$ and $M^B$ converge to zero, 
and $|t_2^d| \to 1$ and $\phi^d \to \pi/2$ for all $d$.
These values correspond to the center of the topological phase with $C = 1$ in the Haldane model, indicated by the star in Fig.~\ref{Fig:haldane}{\bf b}. 
We confirm that different initial conditions converge to the same state (see Supplementary Information). 
Thus, our framework automatically rediscovers the Haldane model with a spontaneous quantum AHE under the condition of maximizing $\sigma_{xy}$. 
The reason why the optimal state is always at the center of the $C=1$ phase is due to the introduction of temperature; at nonzero temperature, $\sigma_{xy}$ becomes largest at the center where the band gap becomes largest in the topological phase. 
We note that the value of $\sigma_{xy}$ in Fig.~\ref{Fig:haldane}{\bf d} is considerably smaller than the quantized value $+1$, which is also due to the finite temperature.

\begin{figure*}[htbp]
        \centering
        \includegraphics[width=18cm,pagebox=cropbox,trim=160 80 170 100, clip]{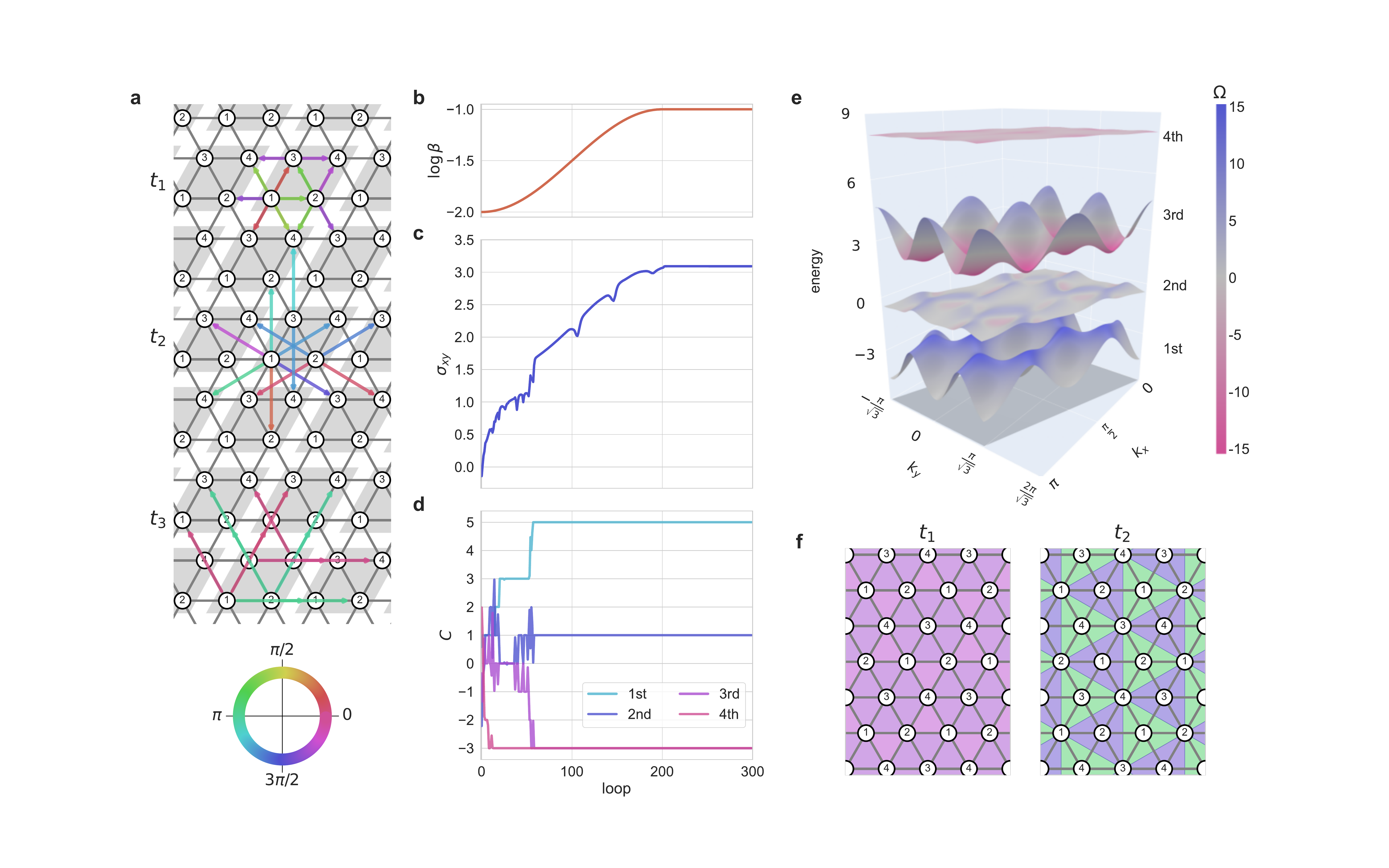} %<left> <lower> <right> <upper>
        \caption{
        {\bf Discovery of a new Hamiltonian showing a six-times larger quantum AHE than the Haldane model.} 
        {\bf a}, A tight-binding model on a triangular lattice with $38$ parameters. 
        The shades denote four-sublattice unit cells. 
        The color of the arrows represents the optimum phase of each hopping, $\phi_m^{ij}$, after the convergence, according to the inset below.
        {\bf b}, Schedule of $\log \beta$. 
        {\bf c, d}, Changes of $\sigma_{xy}$ ({\bf c}) and $C$ for four bands ({\bf d}). 
        $\bf{e}$, The band structure after the convergence, plotted with the Berry curvature $\Omega$. 
        {\bf f}, Fictitious magnetic fluxes defined by the sum of phases along the counter-clockwise direction as $\Phi_m = \sum_{ij} \phi_m^{ij}$ on the smallest triangles by the nearest-neighbor hopping $t_1$ (left) and larger ones by the second-neighbor hopping $t_2$ (right), which are indicated by the same color code as the inset of {\bf a}.
}
        \label{Fig:triangular}
\end{figure*}

Next, we apply our framework to a triangular lattice assuming a four-sublattice unit cell (Fig.~\ref{Fig:triangular}{\bf a}). 
The Hamiltonian consists of the nearest-, second-, and third-neighbor hopping, whose coefficients $t_1^{ij},t_2^{ij},$ and $t_3^{ij}$, respectively, can take different values for different combinations of $i$ and $j$ within the unit cell; 
specifically we take $t_1^{ij} = \exp({\rm i}\phi_1^{ij})$ and  $t_m^{ij} = \sigma(r_m) \exp({\rm i}\phi_m^{ij})$ for $m=2$ and $3$ (see the arrows in Fig.~\ref{Fig:triangular}{\bf a}). 
Thus, the model includes $38$ parameters in total represented by $\bm{\theta} = \{r_2, r_3, \{\phi_1^{ij}\},\{\phi_2^{ij}\}, \{\phi_3^{ij}\} \}$.  
Again, we take $L({\bm \theta})=-\sigma_{xy}({\bm \theta})$ to be minimized, with a schedule of temperature shown in Fig.~\ref{Fig:triangular}{\bf b}.
The fermion density is fixed at half filling at each step of the optimization by using the bisection method.

We find that the Chern numbers for four bands converge to $C=5$, $1$, $-3$, and $-3$ from the lower band, as shown in Fig.~\ref{Fig:triangular}{\bf d}. 
This indicates that $\sigma_{xy}$ reaches $6$ at half filling, which is six times larger than that in the Haldane model, although $\sigma_{xy}$ in Fig.~\ref{Fig:triangular}{\bf c} is much smaller due to the finite temperature similar to the previous case. 
The band structure is shown in Fig.~\ref{Fig:triangular}{\bf e} with the Berry curvature $\Omega$ (see Methods). 
Note that the system recovers (approximately) threefold rotational symmetry after the convergence (see Supplementary Information). 
$\Omega$ of the lowest energy band is positive at all wave numbers, whose sum gives the largest $C=5$, while the other bands include negative contributions. 
This indicates that our framework tries to maximize $C$ for the lowest energy band.
We note that the same conclusion is obtained for many other initial conditions, while some cases converge to $C=3$, $3$, $-1$, and $-5$ from the lower band, which gives the same value of $\sigma_{xy}=6$. 
The reason why the solution in Fig.~\ref{Fig:triangular} is rather preferred is the finite temperature introduced in the optimization process, for the same reason as in the honeycomb lattice model for which the center of the topological phase was obtained (see Supplementary Information).

Let us discuss the optimized parameters. 
We find that both $|t_2|$ and $|t_3|$ converge to $\simeq 1$, while the phases take the various values shown by colors in Fig.~\ref{Fig:triangular}{\bf a}. 
We show, however, that their sums along closed loops, $\Phi_m = \sum \phi_m^{ij}$, representing the fictitious magnetic fluxes, take some regular values: 
$\Phi_1\simeq 7\pi/4$ for the smallest triangles composed of $t_1$, and $\Phi_2$ takes $\simeq 0.91\pi$ and $\simeq 1.59\pi$ for larger triangles of $t_2$ facing right and left, respectively (Fig.~\ref{Fig:triangular}{\bf f}), while $\Phi_3$ is always $\simeq \pi$ ($\phi_3^{ij}$ is either $\simeq 0$ or $\pi$).
Although $\phi_m^{ij}$ take different values for different initial conditions, $\Phi_m$ converge to the same values. 
These results indicate that our framework automatically finds a new model whose complex hoppings realize spontaneous fictitious magnetic fluxes to maximize $\sigma_{xy}$, which is hard to obtain by human intuition. 
Based on this discovery, we can also refine the Hamiltonian by taking more regular values of the phases (multiples of $\pi/4$); 
see Supplementary Information.

\begin{figure*}[htbp]
        \centering
        \includegraphics[width=18cm,pagebox=cropbox,trim=100 120 120 160, clip]{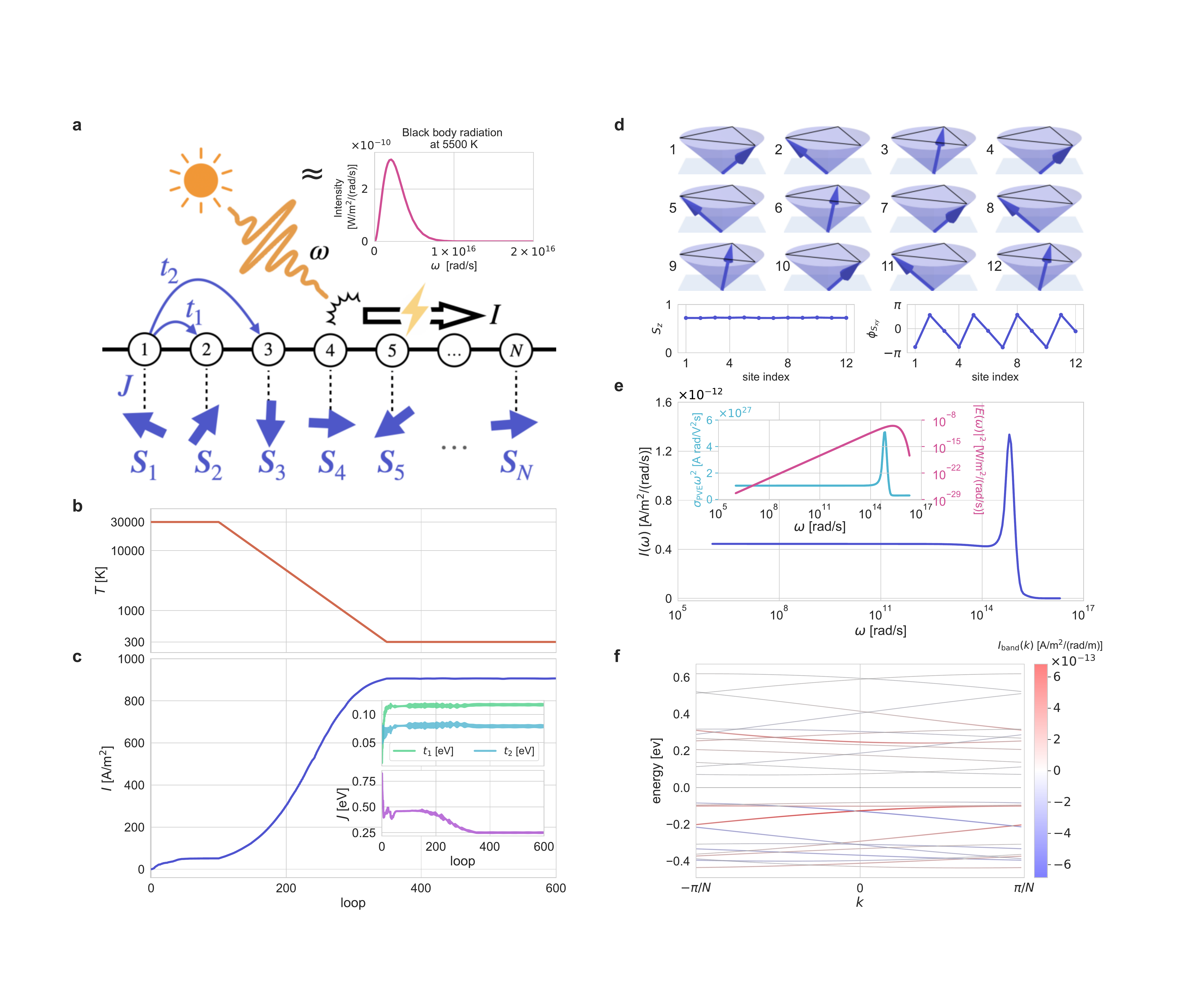} %<left> <lower> <right> <upper>
        \caption{
        {\bf Automatic construction of a Hamiltonian for electrons moving on a noncoplanar spin texture, which can generate $\bm{ \sim 900}$~A/m$\bm{^2}$ under solar radiation.} 
        {\bf a} , Schematic of the system.  
        A photocurrent is generated by solar radiation (blackbody radiation at $5500$~K in the inset) onto the one-dimensional spin-charge coupled system. 
        {\bf b}, Schedule of temperature $T$~[K]. 
        {\bf c}, Change of the photocurrent $I$~[A/m$^2$]. 
        The insets show the changes of $t_1$, $t_2$, and $J$~[eV]. 
        {\bf d}, Spin configurations after the convergence (top), plotted with the $z$ components $S_z$ (left bottom) and the angles of spins projected onto the $xy$ plane, $\phi_{S_{xy}}$ (right bottom). 
        The $S_z$ axis is taken in the direction of the total magnetization. 
        {\bf e}, $\omega$ dependence of $I(\omega) = \sigma_{\rm PVE}(\omega)|E(\omega)|^2$, where $\sigma_{\rm PVE}(\omega)$ is the nonlinear optical conductivity and $|E(\omega)|^2$ is the intensity of solar light (inset).
        {\bf f}, The band structure of electrons. The color represents the contribution to $I$ from each band. 
}
        \label{Fig:photovoltaic}
\end{figure*}

Finally, we apply our framework to optimize the PVE from the so-called shift current in noncentrosymmetric bulk systems~\cite{PhysRev.134.A1313,doi:10.1063/1.1655453,PhysRevB.23.5590,PhysRevLett.109.236601,PhysRevB.99.045121}. 
For simplicity, here we focus on (quasi-)one-dimensional spin-charge coupled systems where the spin configurations break spatial inversion symmetry~\cite{PhysRevB.104.L180407}.
The schematic is shown in Fig.~\ref{Fig:photovoltaic}{\bf a}.
Note that the model approximately describes chiral magnetic metals, such as CrNb$_3$S$_6$~\cite{PhysRevB.92.220412} and Yb(Ni$_{1-x}$Cu$_x$)$_3$Al$_9$~\cite{doi:10.7566/JPSJ.86.124702}. 
The Hamiltonian reads
\begin{eqnarray}
\mathcal{H} &=& \sum_{i,\alpha} 
\left(t_1c_{i \alpha}^{\dagger} c_{i+1\alpha} 
+ t_2c_{i \alpha}^{\dagger} c_{i+2\alpha}
+ {\rm H.c.}  \right) \nonumber \\
& & +J \sum_{i,\alpha, \beta} c_{i\alpha}^\dagger \bm{\sigma}_{\alpha \beta} c_{i\beta} \cdot \bm{S}_i,
\end{eqnarray}
where $c_{i\alpha}^\dagger$ ($c_{i\alpha}$) denotes the creation (annihilation) operator of an electron at site $i$ with spin $\alpha$. 
Here, we take 
$t_1 = \sqrt{2}{\rm tanh}(r_t) \cos(\theta_t) \times 0.1$~[eV], 
$t_2 = \sqrt{2}{\rm tanh}(r_t) \sin(\theta_t) \times 0.1$~[eV], and 
$J = \log(1+\exp(r_J))$~[eV];
the spins are treated as classical and their configurations are parametrized as 
${\bm S}_i = (\sin \theta_i \cos \phi_i,  \sin \theta_i \sin \phi_i,  \cos \theta_i)$, 
with $\theta_i = \pi \sigma(\eta_i)$. 
We set the number of sublattice sites to $N=12$. 
Thus, the model includes $3+2N=25$ parameters in total represented by ${\bm \theta} = \{r_t, \theta_t, r_J, \{\eta_i\}, \{\phi_i\} \}$. 
The quantity of our interest is the photocurrent under solar radiation, defined as $I = \int d\omega \sigma_{\rm PVE}(\omega) |E(\omega)|^2 $~[A/m$^2$], where $\sigma_{\rm PVE}(\omega)$ is the nonlinear optical conductivity, and $|E(\omega)|^2$ denotes the intensity of the linearly polarized solar light with frequency $\omega$, approximately given by blackbody radiation at $T = 5500$~K (the inset of Fig.~\ref{Fig:photovoltaic}{\bf a}) (see Methods); we take $L({\bm \theta})=-I$. 
We consider a three-dimensional system in which the one-dimensional chains are arranged in a square lattice fashion for simplicity, taking the lattice constants $a_z = 9$~$\mathrm{\mathring{A}}$ in the chain direction and $a_x = a_y = 4$~$\mathrm{\mathring{A}}$ in the orthogonal directions, referring to a chiral magnet~\cite{doi:10.7566/JPSJ.86.124702}.
The fermion density is fixed at half filling as for the previous model.

Figure~\ref{Fig:photovoltaic}{\bf c} shows the optimization process of the photocurrent $I$ under the schedule of temperature shown in Fig.~\ref{Fig:photovoltaic}{\bf b}. 
We obtain $I \sim 900$~A/m$^2$ after the convergence.  
This value is comparable or larger than those for Ge semiconductors~\cite{SINGH201236} and perovskites substances~\cite{doi:10.1021/jz500113x,doi:10.1021/acsanm.0c00888}. 
Changes of the parameters $t_1$, $t_2$, and $J$ are plotted in the inset of Fig.~\ref{Fig:photovoltaic}{\bf c}.
The optimized spin configuration is an umbrella-shaped chiral state with three-site period, as shown in Fig.~\ref{Fig:photovoltaic}{\bf d}.  
We also note that spin configurations with four-site period are also obtained for different initial conditions, but they generate smaller $I$ (see Supplementary Information).

To elaborate the mechanism behind the optimization of the photocurrent, we plot the $\omega$ dependence of $I(\omega) = \sigma_{\rm PVE}(\omega) |E(\omega)|^2 $ in Fig.~\ref{Fig:photovoltaic}{\bf e}, together with $\sigma_{\rm PVE}(\omega) \omega^2$ and $|E(\omega)|^2$ in the inset.
We find that $I(\omega)$ has a sharp peak at $\omega \sim 7.15 \times 10^{14}$~[rad/s], due to the peak of $\sigma_{\rm PVE}(\omega)\omega^2$ locating at the frequency where $|E(\omega)|^2$ becomes large.
We show that dominant contributions to the peak come from the interband processes between the conduction and valence bands split by $2J \simeq 0.5$~[eV]~$\simeq 7.15 \times 10^{14}$~[rad/s], as shown in Fig.~\ref{Fig:photovoltaic}{\bf f} (see Methods). 
The results indicate that the enhanced photocurrent of $\sim 900$~A/m$^2$ under solar radiation is generated by the band engineering with automatic optimization of $t_1$, $t_2$, $J$, and the spin configurations. 
We note that the peak value of $\sigma_{\rm PVE}(\omega) \sim 0.04$~A/V$^2$ is considerably large compared to existing materials, such as BaTiO$_3$~\cite{PhysRev.134.A1313,PhysRevLett.109.116601} and TaAs~\cite{Osterhoudt2019}, and is also even an order of magnitude larger than the value obtained in the previous theoretical study~\cite{PhysRevB.104.L180407}, while we may need substantially-large competing magnetic interactions to stabilize the umbrella spin configuration at room temperature.

Through the applications to AHE and PVE, our framework has proven capable of automatically finding Hamiltonians that optimize the physical properties of interest. 
The key aspect is in the use of automatic differentiation in the inverse problem, which provides the derivatives of the objective function in terms of large number of parameters; 
although the current studies are limited to several tens of parameters, we can practically deal with a million or more. 
Since automatic differentiation is a versatile technique, our framework has a wide range of the applicability, 
such as first-principles Hamiltonians computed by the Kohn-Sham equations, strongly correlated electron systems, quantum spin systems, and interacting bosonic systems. 
In addition, it is applicable to a wide range of physical properties to be optimized. 
%In particular, it is promising to design Hamiltonians with physical constraints by incorporating constrained optimization methods.
Thus, our finding will open up new directions to explore new models and principles in materials science.

\section*{Methods}
\label{sec:method}

\subsection*{Application to the AHE}
\label{subsec-method-topological}

The Hall conductivity is calculated by using the Kubo formula as 
\begin{equation}
\sigma_{xy} = -\frac{e^2}{h}\frac{V}{2\pi N_{\bm k}} \sum_{m,n,\bm{k}} \bigl(f(E_{\bm{k}n},\beta)-f(E_{\bm{k}m},\beta)\bigr) \Omega(\bm{k}), 
\label{sigmaxy}
\end{equation}
where $e$ is the elementary charge, $h$ is the Planck constant, $V$ is the volume of the Brillouin zone, $N_{\bm k}$ is the number of $\bm k$ points, $f(E,\beta)$ is the Fermi distribution function at inverse temperature $\beta$, $E_{\bm{k}n}$ is the energy at $\bm{k}$ in $n{\rm th}$ band; $\Omega(\bm{k})$ is the Berry curvature given by
\begin{equation}
\Omega(\bm{k}) = {\rm Im} \frac{
    \langle \bm{k}n | \frac{\partial \mathcal{H}}{\partial k_y} | \bm{k}m \rangle 
    \langle \bm{k}m | \frac{\partial \mathcal{H}}{\partial k_x}| \bm{k}n \rangle
    }
{(E_{\bm{k}n}-E_{\bm{k}m})^2 + {\rm i}\delta}, 
\label{eq:omega}
\end{equation}
where $|{\bm k}n\rangle$ is an eigenstate at $\bm k$ in $n$th band. 
We take $e=h=1$, $N_{\bm k}=100^2$, and $\delta = 10^{-5}$.

The optimization starts from initial parameters randomly chosen as 
$M^A, M^B \in (-1,1)$, $r^a \in (0,1)$, and $\phi^a \in (-\pi, \pi)$ 
for the honeycomb lattice model, and
$r_2,r_3 \in (0,1)$ and $\phi_1^{ij}, \phi_2^{ij}, \phi_3^{ij} \in (-\pi, \pi)$  
for the triangular lattice model. 
Automatic differentiation is implemented using JAX~\cite{jax2018github}. 
Note that $\frac{\partial \mathcal{H}}{\partial k_x}$ and $\frac{\partial \mathcal{H}}{\partial k_y}$ in Eq.~\eqref{eq:omega} are also calculated by using automatic differentiation. 
We employ RMSPROP~\cite{hinton2012neural} as an optimization method, in which we take the learning rate, the decay factor, and the infinitesimal as $0.1$, $0.99$, and $10^{-8}$, respectively.

\subsection*{Application to the PVE} 
\label{subsec-method-pve}

According to the second-order optical response theory~\cite{PhysRevB.99.045121,PhysRevB.104.L180407}, a nonlinear electric current produced by electric fields $E(\omega_1)$ and $E(\omega_2)$ with two frequencies $\omega_1$ and $\omega_2$, respectively, is given by
\begin{eqnarray}
I(\omega_1 + \omega_2; \omega_1, \omega_2) &=& \nonumber \\ 
\sigma_{\rm opt}(\omega_1 &+& \omega_2; \omega_1, \omega_2) E(\omega_1) E(\omega_2),
\end{eqnarray}
with the second-order optical conductivity $\sigma_{\rm opt}(\omega_1 + \omega_2; \omega_1, \omega_2)$. 
In the case of $\omega_1 = -\omega_2$, a DC current is generated as 
\begin{eqnarray}
I(\omega) = \sigma_{\rm PVE}(\omega) |E(\omega)|^2, 
\label{eq:Iomega}
\end{eqnarray}
where $I(\omega) = I(0; \omega, -\omega)$ and  $\sigma_{\rm PVE}(\omega) = \sigma_{\rm opt}(0; \omega, -\omega)$. 
The $\omega$ integral $I=\int d\omega I(\omega)$ gives a photocurrent generated by the shift current mechanism~\cite{PhysRevB.23.5590,PhysRevLett.109.236601,PhysRevB.104.L180407}, which is used for the objective function in the main text. 
We approximate solar radiation by blackbody radiation $B(\omega, T)$ at $5500$~K as 
\begin{equation}
|E(\omega)|^2 = 2\mu_0 c C_{\rm solar} \frac{B(\omega, T=5500\,{\rm  K})}{\int d\omega B(\omega , T=5500\,{\rm K}) }, 
\end{equation}
where $\mu_0$, $c$, and $C_{\rm solar}$ are the magnetic constant, speed of light, and solar constant, respectively; 
\begin{equation}
B(\omega, T) = \frac{\hbar \omega^3}{4\pi^3 c^2} \frac{1}{\exp (\frac{\hbar \omega}{k_B T})-1}, 
\end{equation}
where $\hbar$ and $k_B$ are the reduced Planck constant and the Boltzmann constant, respectively. 
In Eq.~\eqref{eq:Iomega}, $\sigma_{\rm PVE}(\omega)$ is computed as~\cite{PhysRevB.104.L180407} 
\begin{widetext}
\begin{eqnarray}
\sigma_{\rm PVE}(\omega) = - \frac{V}{(2\pi)^3} \frac{1}{N_k \omega^2} (\sigma_{\rm PVE,1} + \sigma_{\rm PVE,2} + \sigma_{\rm PVE,3} + \sigma_{\rm PVE,4}), 
\end{eqnarray}
where 
\begin{eqnarray}
\sigma_{\rm PVE,1} &=& \sum_{k,a} f(E_k,\beta) J_{aa}^{(3)} \label{eq:pve_1},\\
\sigma_{\rm PVE,2} &=& \sum_{k,a,b} \biggl(
\frac{f_{ab} J_{ab}^{(1)} J_{ba}^{(2)}}{ \omega + {\rm i}\gamma - E_{ab}} + 
\frac{f_{ab} J_{ab}^{(1)} J_{ba}^{(2)}}{-\omega + {\rm i}\gamma - E_{ab}} \biggl) \label{eq:pve_2},\\
\sigma_{\rm PVE,3} &=& \sum_{k,a,b} \frac{f_{ab} J_{ab}^{(2)} J_{ba}^{(1)}}{{\rm i}\gamma - E_{ab}} \label{eq:pve_3},\\
\sigma_{\rm PVE,4} &=& \sum_{k,a,b,c} 
\frac{J_{ab}^{(1)} J_{bc}^{(1)} J_{ca}^{(1)}}{{\rm i}\gamma - E_{ca}} \biggl( 
    \frac{f_{ab}}{ \omega + {\rm i}\gamma -E_{ba}} +     
    \frac{f_{ca}}{ \omega + {\rm i}\gamma -E_{cb}} +     
    \frac{f_{ab}}{-\omega + {\rm i}\gamma -E_{ba}} +     
    \frac{f_{cb}}{-\omega + {\rm i}\gamma -E_{cb}}  \biggr).
\label{eq:pve_4}
\end{eqnarray}
\end{widetext}
Here, $a$, $b$, and $c$ denote the bands; 
$E_{ab} = E_{ka} - E_{kb}$, $f_{ab} = f(E_{ka},\beta) - f(E_{kb},\beta)$, and $J_{ab}^{(n)} = \langle ka | \frac{\partial^n \mathcal{H}}{\partial k^n} | kb \rangle$. 
We use $V = \frac{(2 \pi)^3}{a_x a_y N a_z}$, $N_k = 100$, and $\gamma=2\pi\times10^{13}$~[rad/s]. 
$\frac{\partial^n\mathcal{H}}{\partial k^n}$ in $J_{ab}^{(n)}$ are calculated by using automatic differentiation.
We also calculate the contribution to $I$ from each $k$ point in each band, $I_{\rm band}(k)$, by calculating $I$ without taking the summations of $k$ and the band indices in Eqs.~\eqref{eq:pve_1}--\eqref{eq:pve_4}. 
The optimization starts from initial parameters randomly chosen as $r_t \in (-1,1)$, $\theta_t \in (-\pi, \pi)$, $r_J \in (0, 0.5)$, 
$\eta_i \in (-1,1)$, and  $\phi_i \in (-\pi, \pi)$.

%\section*{Data availability}
%\label{sec:data}
%All the data can be generated from the code below.

\section*{Code availability}
\label{sec:code}
We have published the code to reproduce all the results on https://github.com/koji-inui/automatic-hamiltonian-design.git.

\section*{Acknowledgments}
The authors thank Y. Kato, S. Okumura, R. Pohle, and K. Shimizu for fruitful discussions.
This work is supported by Japan Society for the Promotion of Science (JSPS) KAKENHI Grant Nos.~JP19H05825 and 20H00122, and JST CREST Grant No.~JPMJCR18T2.

%\section*{Author information}
%\label{sec:author}
%
%
%\subsection*{Contribution}
%\label{sec:contribution}
%K.I. conceived and implemented the algorithm through the discussion with Y.M.
%K.I. and Y.M. conceived the models, interpreted the results, and wrote the manuscript. 
%
%
%\section*{Ethics declarations}
%\label{sec:ethics}
%
%\subsection*{Competing interests}
%\label{sec:Competing interests}
%K.I. has filed a patent based on the algorithm reported in this paper. 

%\bibliographystyle{unsrtnat} % これでnature のフォーマットになる。
\bibliography{main}% Produces the bibliography via BibTeX.

\section*{Supplementary information}
\label{sec:supplement}

\setcounter{equation}{0}
\setcounter{figure}{0}
%\documentclass[
% aps,
% amsmath,amssymb,
%]{revtex4-2}
%
%\usepackage{dcolumn}% Align table columns on decimal point
%\usepackage{bm}% bold math
%\usepackage[dvipdfmx]{graphicx,color} % for using Japanese in platex
%%\usepackage[pdftex]{graphicx,color} % for publishing by pdflatex
%\usepackage{comment}
%\usepackage{natbib}
%\usepackage{breakcites}
%
%\usepackage{here}
%\usepackage[normalem]{ulem}
%\usepackage[breaklinks=true]{hyperref}
%%\usepackage{hyperref}% add hypertext capabilities
%\usepackage{braket}
%\usepackage{url}
%\usepackage{booktabs}
%\usepackage{array}
\newcolumntype{M}[1]{>{\centering\arraybackslash}m{#1}}
\renewcommand{\thesection}{S\arabic{section}}
\renewcommand{\thetable}{S\arabic{table}}
\renewcommand{\thefigure}{S\arabic{figure}}

%\begin{document}

\section{Rediscovery of the Haldane Model: initial condition dependence}
\label{supp_sec:haldane}

\begin{figure*}[htbp]
        \centering
        \includegraphics[width=15cm,pagebox=cropbox,trim=0 0 0 0, clip]{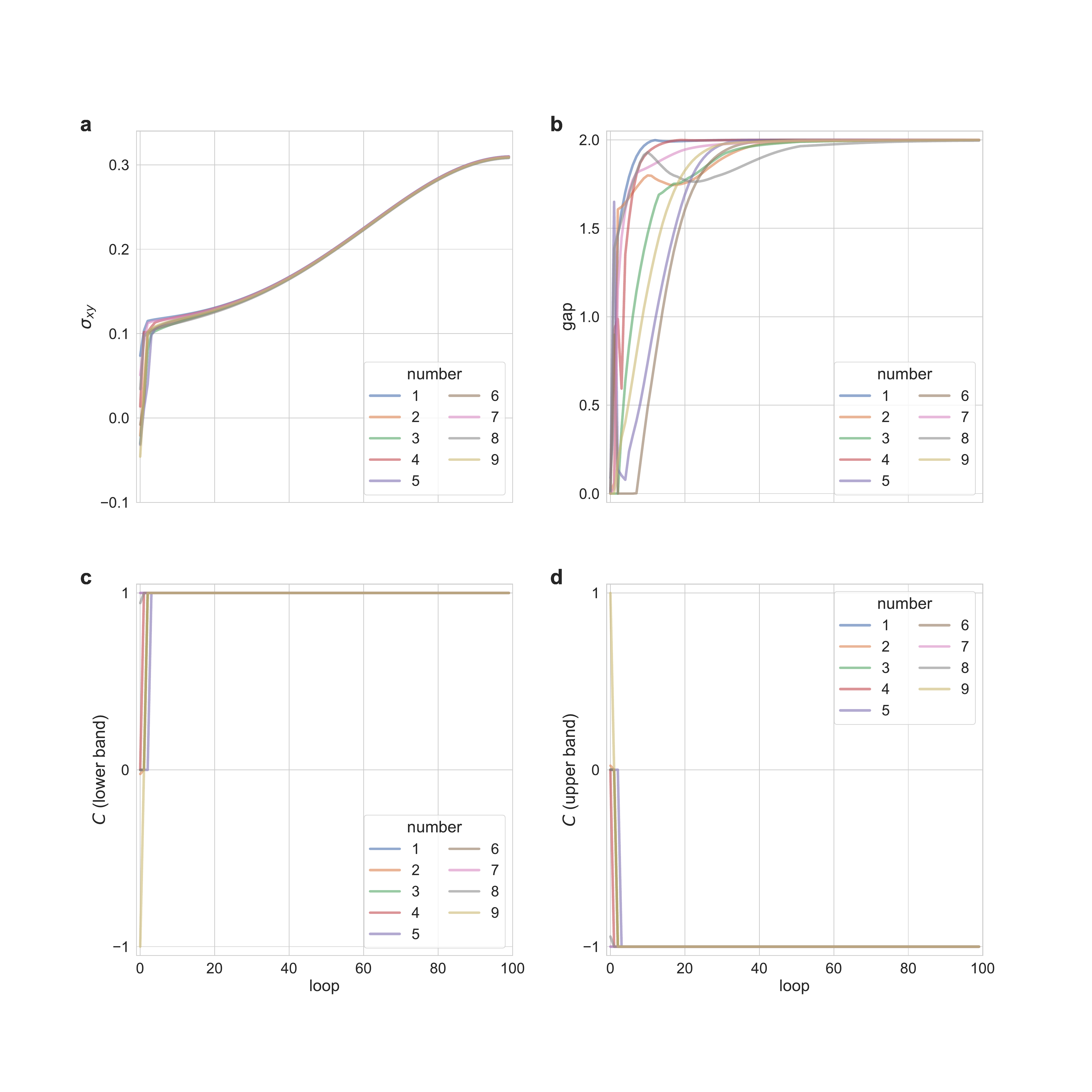} %<left> <lower> <right> <upper>
        \caption{
        Changes of the Hall conductivity $\sigma_{xy}$ ({\bf a}), the band gap ({\bf b}), the Chern numbers of the lower band ({\bf c}) and the upper band ({\bf d}) for the tight-binding model on a honeycomb lattice in Eq.~(1) in the main text. 
        The number labels nine different sets of initial parameters. 
}
        \label{Fig:haldane_init}
\end{figure*}

\begin{figure*}[htbp]
        \centering
        \includegraphics[width=14cm,pagebox=cropbox,trim=0 200 0 300, clip]{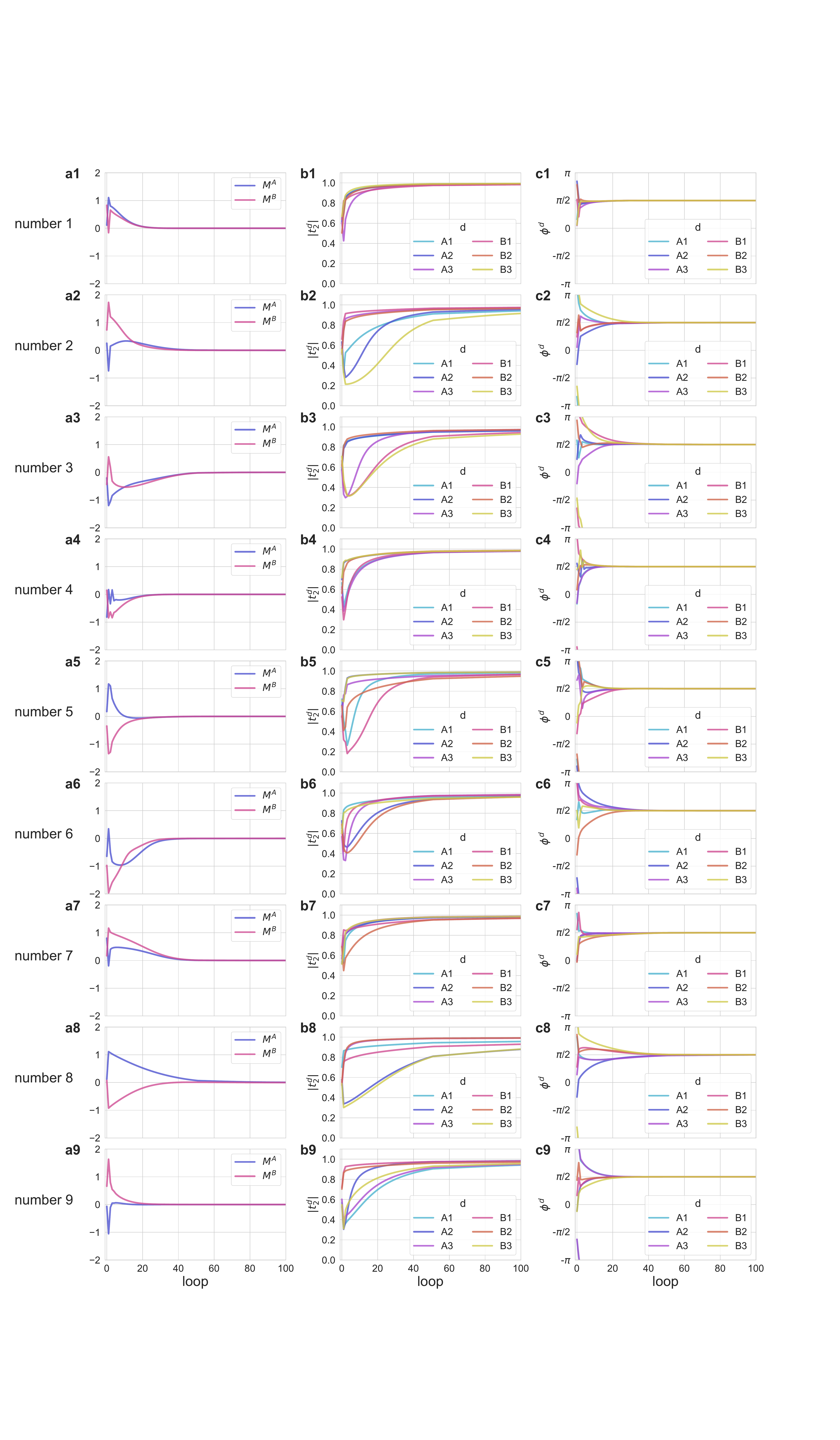} %<left> <lower> <right> <upper>
        \caption{
        Changes of $M^a$ $(\bf{a1\mathchar`-9})$, $|t_2^d|$ $(\bf{b1\mathchar`-9})$ and $\phi^d$ $(\bf{c1\mathchar`-9})$  for the tight-binding model on a honeycomb lattice in Eq.~(1) in the main text for the nine sets of initial parameters. 
}
        \label{Fig:haldane_param}
\end{figure*}

Figure~\ref{Fig:haldane_init} shows the optimization process of $\sigma_{xy}$, the band gap, and the Chern number of each band for nine sets of different initial parameters of the Hamitonian in Eq.~(1) in the main text. 
All the cases converge to the same results. 
The changes of the model parameters are shown in Fig.~\ref{Fig:haldane_param}. 
All of them converge to the same results corresponding to the center of the $C=1$ phase of the Haldane model indicated by the star in Fig.~2{\bf b} in the main text.

\section{Discovery of a new Hamiltonian on a triangular lattice}
\label{supp_sec:triangular}

\subsection{Initial condition dependence}
\label{supp_subsec:triangular_initial}

\begin{figure*}[htbp]
        \centering
        \includegraphics[width=10cm,pagebox=cropbox,trim=0 0 0 0, clip]{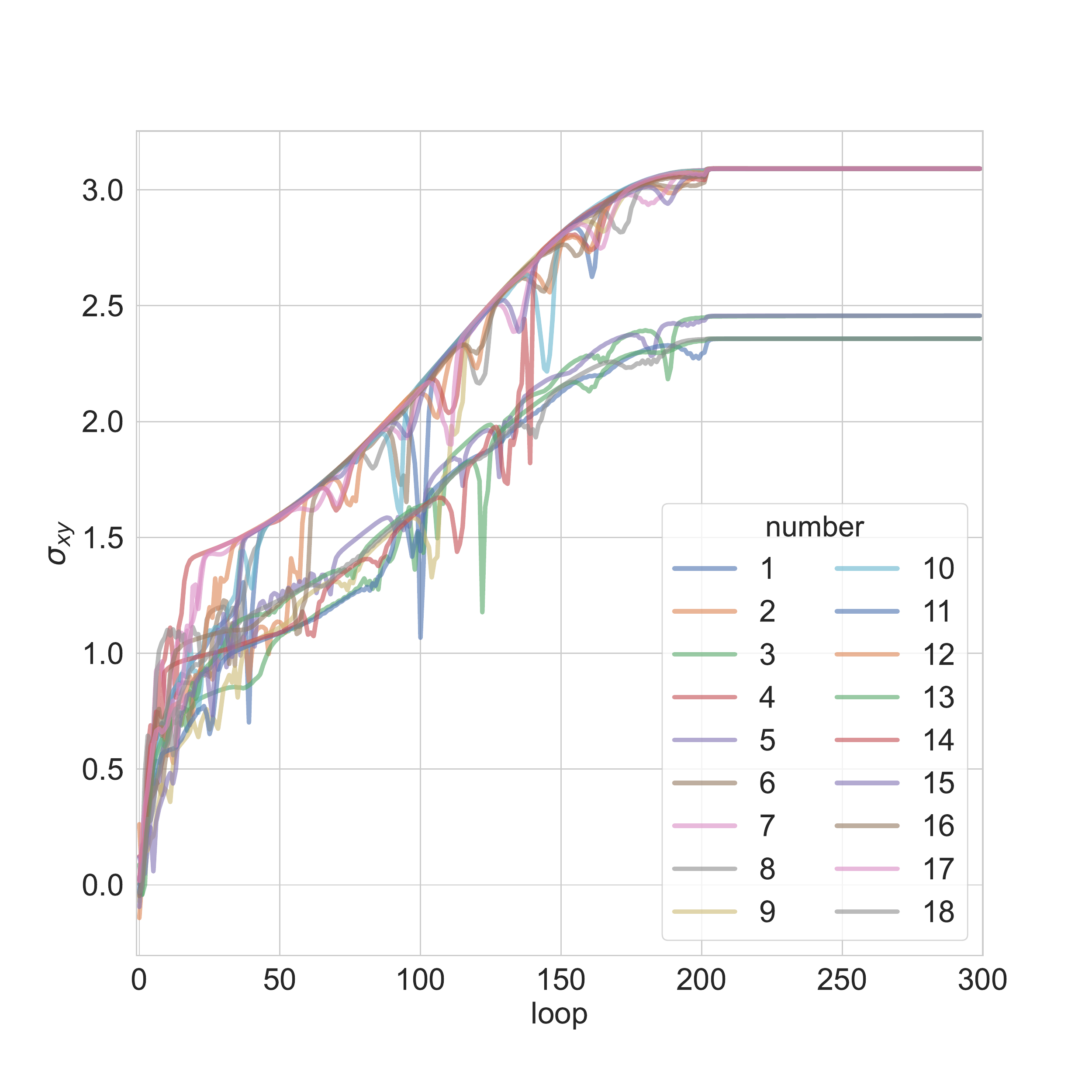} %<left> <lower> <right> <upper>
        \caption{
        Changes of the Hall conductivity $\sigma_{xy}$ for the tight-binding model on a triangular lattice in the main text. 
        The number labels eighteen different sets of initial parameters. 
}
        \label{Fig:triangular_init}
\end{figure*}

Figure~\ref{Fig:triangular_init} shows the optimization process of $\sigma_{xy}$ for eighteen sets of different initial parameters of the Hamitonian on a triangular lattice in the main text. 
$\sigma_{xy}$ converges to three different values; the largest one takes the value around $3.2$, while the rest two are around $2.5$ and $2.4$.

\begin{table*}[htbp]
  \centering
  \renewcommand{\arraystretch}{1.2}
  \begin{tabular}{|c|c|c|c|c|}
    \hline
    \multicolumn{4}{|c|}{$C$} &  Number of \\
    1st & 2nd & 3rd & 4th  &  initial conditions  \\
    \hline 
    5 & 1 & -3 & -3  & 10  \\
    \hline
    3 & 3 & -1 & -5  & 3  \\
    \hline
    3 & 2 & -2 & -3  & 2  \\
    \hline
    3 & 1 & -1 & -3  & 3  \\
    \hline
  \end{tabular}
  \caption{
  The Chern numbers of the four bands after the convergence starting from the eighteen different sets of initial parameters. 
  The number of the initial conditions that converge to each result is also shown.
  }
  \label{table:triangular}
\end{table*}

Table~\ref{table:triangular} shows the Chern numbers of the four bands obtained from the different initial parameters. 
The number of the initial conditions that converge to the corresponding result is also listed. 
The majority is the group with $C=5$, $1$, $-3$, and $-3$, which gives $\sigma_{xy} \simeq 3.2$ in Fig.~\ref{Fig:triangular_init}: 
ten out of eighteen initial conditions converge to this result, and one of them is discussed in the main text. 
Meanwhile, the group with $C=3$, $3$, $-1$, and $-5$ (three out of eighteen) also gives $\sigma_{xy} \simeq 3.2$ as the sum of $C$ for the filled two bands are the same as the majority group. 
The rest two groups give the smaller $\sigma_{xy} \simeq 2.5$ and $2.4$ in Fig.~\ref{Fig:triangular_init}. 
Note that the four groups give $\sigma_{xy}=6$, $6$, $5$, and $4$ at zero temperature, corresponding to the sum of $C$ for the filled two bands.

\subsection{Band structure}
\label{supp_subsec:triangular_symmetry}

\begin{figure*}[htbp]
        \centering
        \includegraphics[width=14cm,pagebox=cropbox,trim=0 50 0 50, clip]{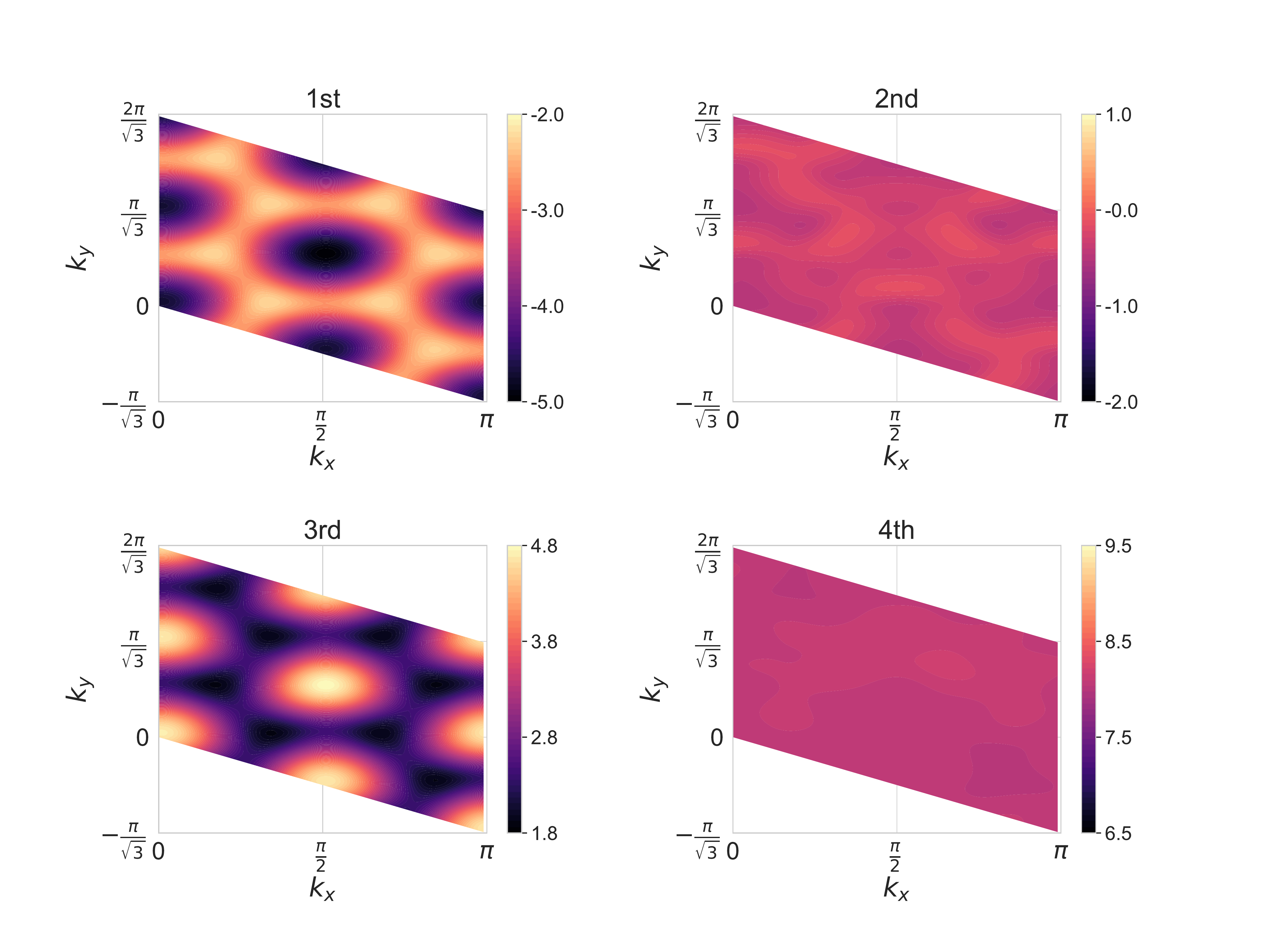} %<left> <lower> <right> <upper>
        \caption{
        The band structure after the convergence for the tight-binding model on a triangular lattice in the main text. 
        All the bands restore approximately threefold rotational symmetry, while it is not so obvious for the 2nd and 4th bands since they are almost flat. 
}
        \label{Fig:bandstructure}
\end{figure*}

Figure~\ref{Fig:bandstructure} shows the energy dispersion of each band for one of the optimal solution shown in Fig.~3{\bf e} in the main text.  
The results indicate that the threefold rotational symmetry is approximately restored by the automatic optimization of $\sigma_{xy}$.

\subsection{Refinement of the model parameters}
\label{supp_subsec:triangular_simplified}

\begin{figure*}[htbp]
        \centering
        \includegraphics[width=14cm,pagebox=cropbox,trim=0 0 0 0, clip]{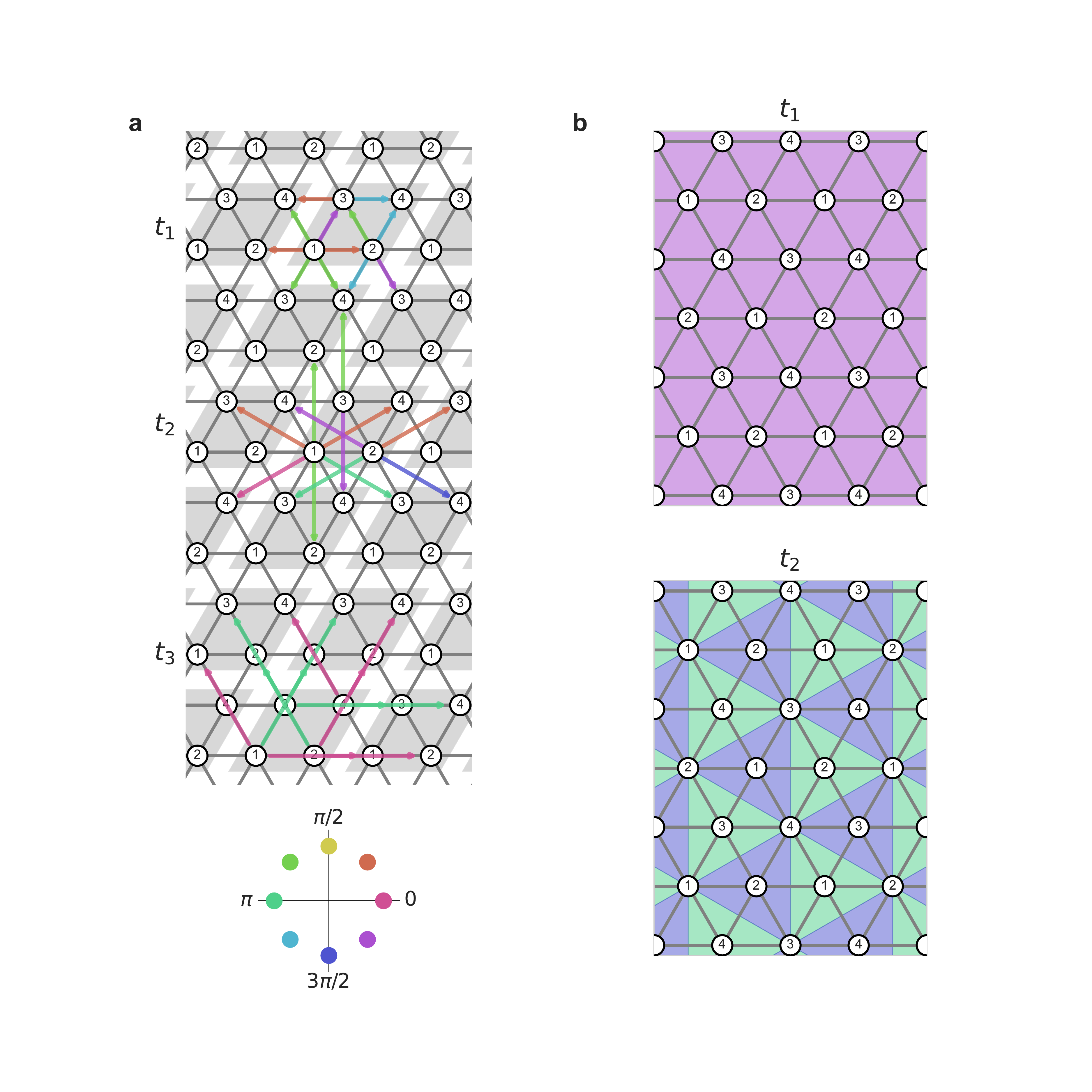} %<left> <lower> <right> <upper>
        \caption{
        $\bf{a}$, The phases of the hopping coefficients of the tight-binding Hamiltonian on a triangular lattice after the refinement.
        All the phases are set to be multiples of $\pi/4$, as represented by the color of the arrows.
        The amplitudes of the hopping are taken as $|t_1|=|t_2|=|t_3|=1$. 
        $\bf{b}$, Sum of the phases $\Phi_m$ defined in the main text for the triangles composed of $t_1$ (top) and $t_2$ (bottom). 
}
        \label{Fig:triangular_phase}
\end{figure*}

As discussed in the main text, the phases of the hopping coefficients converge to rather scattered values. 
We find, however, that it is possible to regularize the parameters by hand while keeping the value of $\sigma_{xy}$ by the following procedures. 
First, we set $|t_1| = |t_2| = |t_3| =1$ and transform the hopping coefficients so as to shift the band bottom of the 1st band around $\bm{k}_0 = (\pi/2, \pi/(2\sqrt{3}))$ to the $\Gamma$ point as $\tilde{t}^{ij}_m= t^{ij}_m \exp({\rm i} \bm{d}^{ij}\cdot \bm{k}_0)$, where $\bm{d}^{ij}$ is a vector from site $i$ to $j$ in real space. 
Next, we set the phase of each hopping to a multiple of $\pi/4$ close to the value after the transformation. 
In addition, we tune a few of them by hand so as to make $\Phi_m$ spatially uniform. 
Then, we obtain the phases shown in Fig.~\ref{Fig:triangular_phase}{\bf a}.
These procedures provide us a Hamiltonian with regularized hopping coefficients, which looks similar to Fig.~3{\bf f} in the main text but has spatially uniform fictitious magnetic fluxes as shown in Fig.~\ref{Fig:triangular_phase}{\bf b}. 
This demonstrates that our framework is useful for constructing a new Hamiltonian in a regularized form. 

\section{Application to the photovoltaic effect: initial condition dependence}
\label{supp_sec:pve}

\begin{figure*}[htbp]
        \centering
        \includegraphics[width=10cm,pagebox=cropbox,trim=0 20 0 70, clip]{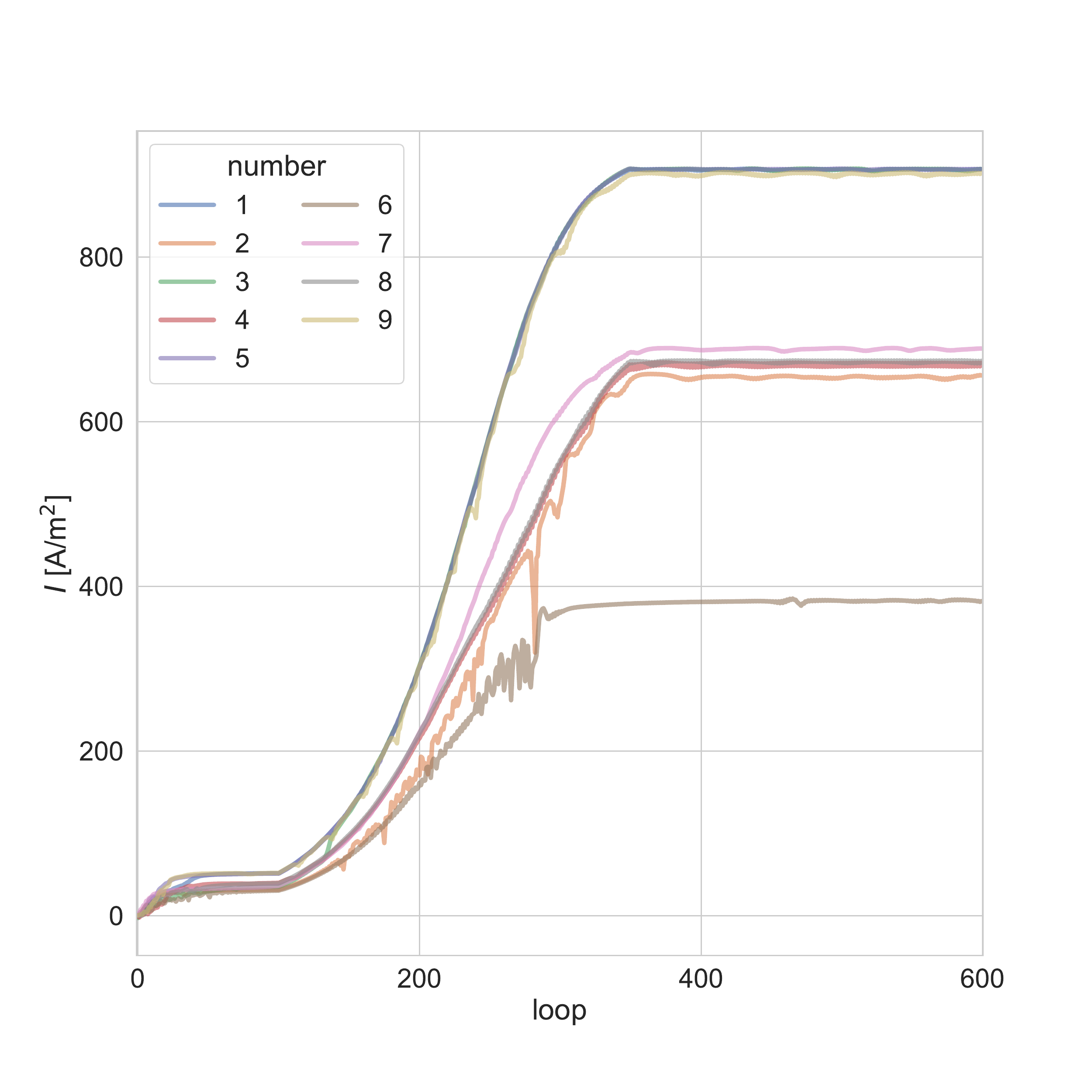} %<left> <lower> <right> <upper>
        \caption{
        Changes of the photocurrent through the optimization process for nine sets of different initial parameters. 
}
        \label{Fig:pve_current}
\end{figure*}

Figure~\ref{Fig:pve_current} shows the optimization process of the photocurrent $I$ for nine sets of different initial parameters of the spin-charge coupled model in Eq.~(2) in the main text. 
After the convergence, the results are grouped roughly into three;
one of them gives the largest $I \sim 900$~$\rm{A/m^2}$, as discussed in the main text.

\begin{table*}[htbp]
        \centering
        \centering
        \renewcommand{\arraystretch}{1.2}
        \begin{tabular}{|M{2cm}|M{2cm}|M{2cm}|M{2cm}|M{2cm}|}
          \hline
          number & $I$ [A/m$^2$]  & $t_1$ [eV]  & $t_2$ [eV]  & $J$ [eV] \\
          \hline {\bf 1}   & {\bf 0.247} & {\bf 0.116} & {\bf 0.081}  & {\bf 906} \\
          \hline 2         &  0.239      & -0.118      & 0.078        & 656       \\
          \hline {\bf 3}   & {\bf 0.247} & {\bf 0.116} & {\bf 0.081}  & {\bf 906} \\
          \hline 4         &  0.288      &  -0.122     & -0.072       & 671       \\
          \hline {\bf 5}   & {\bf 0.246} & {\bf 0.116} & {\bf -0.081} & {\bf 906} \\
          \hline 6         &  0.482      & 0.065       & -0.126       & 382       \\
          \hline 7         &  0.256      & 0.114       & 0.084        & 688       \\
          \hline 8         &  0.288      & 0.12        & 0.074        & 673       \\
          \hline {\bf 9}   & {\bf 0.253} & {\bf -0.12} & {\bf 0.074}  & {\bf 902} \\
          \hline
        \end{tabular}
        \caption{
        The values of $I$, $t_1$, $t_2$, and $J$ for the different initial parameters after the convergence. 
        The bold numbers indicate the results categorized into the optimal solution.
}
        \label{table:pve}
\end{table*}

\begin{figure*}[htbp]
        \centering
        \includegraphics[width=9cm,pagebox=cropbox,trim=0 0 0 0, clip]{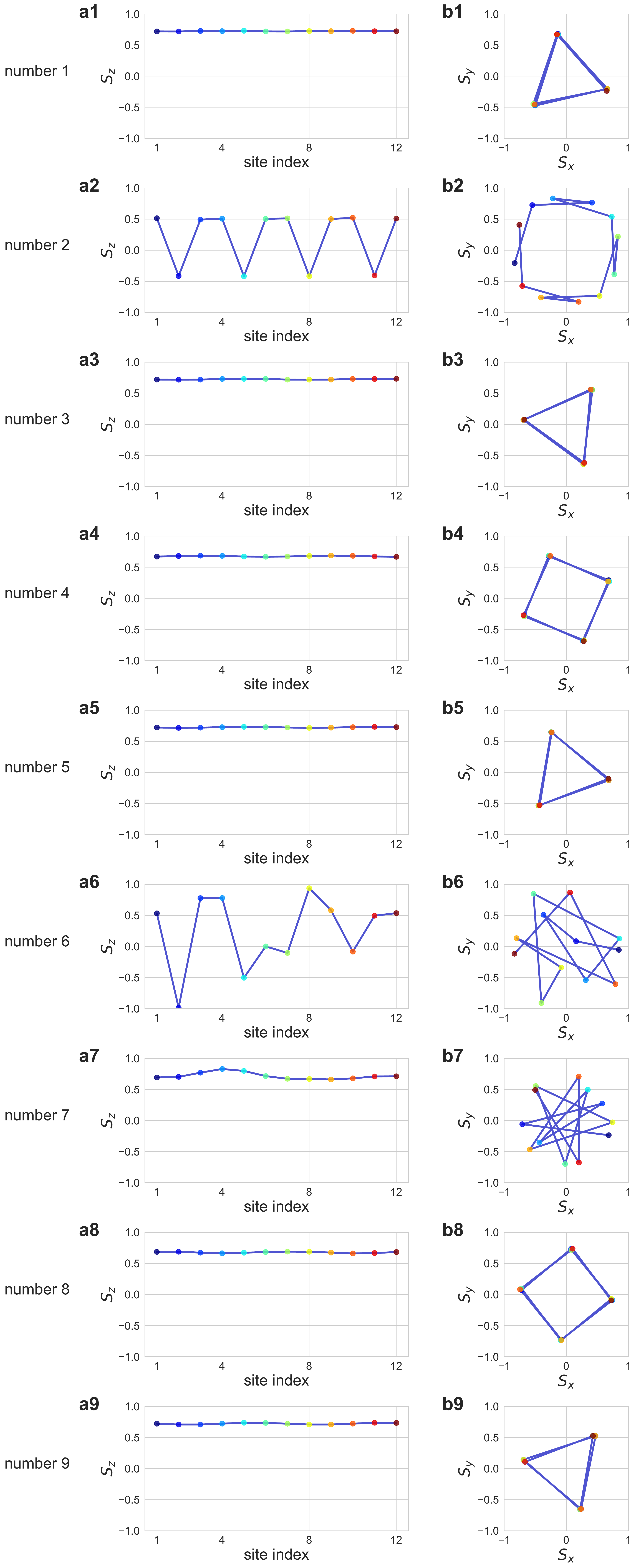} %<left> <lower> <right> <upper>
        \caption{
        Spin configurations for the different initial parameters:
        the $z$ component, $S_z$ ($\bf{a1\mathchar`-9}$), and the $xy$ components, $S_x$ and $S_y$ ($\bf{b1\mathchar`-9}$).
        The $S_z$ axis is taken in the direction of the total magnetization.
        The symbol color represents the site index.
}
        \label{Fig:spins}
\end{figure*}

Table~\ref{table:pve} shows the values of $I$, $t_1$, $t_2$, and $J$ after the convergence starting from the different initial parameters.
Four out of nine solutions converge to the largest value of $I\sim 900$~$\rm{A/m^2}$. 
Spin configurations after the convergence are shown in Fig.~\ref{Fig:spins}.
The results with the optimal value of $I$ (number $1$, $3$, $5$, and $9$) have an umbrella-shaped structure with three-site period.
We note that two of the other results (number $4$ and $8$) gives the modest better value of $I\sim 670$~$\rm{A/m^2}$ with a different umbrella structure with four-site period.
The others have more disordered spin configurations with modulations of $S_z$.
%\end{document}
\end{document}